\documentclass[conference]{IEEEtran}
\IEEEoverridecommandlockouts

\usepackage{amsmath}
\usepackage{amsthm}
\usepackage{amssymb}
\usepackage{mathtools}
\usepackage{booktabs}
\usepackage{graphicx}
\usepackage{xcolor}
\usepackage{framed}
\usepackage{xspace}
\usepackage{url}
\usepackage{algorithm}
\usepackage{algpseudocode}
\usepackage{enumitem}
\usepackage{tikz}
\usepackage{circledtext}
\usepackage{pifont} 
\usetikzlibrary{positioning,arrows.meta,shapes.geometric,fit,calc,backgrounds}

\providecommand{\Description}[1]{}

\newcommand{\rb}{\rule[-0.14ex]{0.42em}{0.82ex}\kern0.13em}

\makeatletter
\renewcommand{\paragraph}[1]{\par\vspace{3pt}\noindent\textbf{#1}\hspace{0.55em}\ignorespaces}
\makeatother

\newtheorem{assumption}{Assumption}

\theoremstyle{remark}

\definecolor{formalshade}{rgb}{0.95,0.95,0.97}
\definecolor{darkblue}{rgb}{0.14,0.22,0.52}
\definecolor{darkorange}{rgb}{1.0,0.549,0.0}
\definecolor{darkgreen}{rgb}{0.0,0.549,0.0}

\newenvironment{assumptionbox}{
\small

\MakeFramed{\advance\hsize-\width\FrameRestore}}
{\endMakeFramed}

\newenvironment{findingbox}{
\small

\MakeFramed{\advance\hsize-\width\FrameRestore}}
{\endMakeFramed}

\newcommand{\V}{\mathcal{V}}

\newcommand{\Pool}{\mathcal{P}}

\newcommand{\Hit}{\textsc{Hit}}
\newcommand{\Miss}{\textsc{Miss}}
\newcommand{\Flood}{\textsc{Flood}}
\newcommand{\Prove}{\textsc{Prove}}
\newcommand{\Profile}{\textsc{Profile}}
\newcommand{\LatencyTest}{\textsc{LatencyTest}}

\newcommand{\Determine}{\textsc{Classify}}
\newcommand{\Classify}{\textsc{Classify}}
\newcommand{\Build}{\textsc{BuildGraph}}

\newcommand{\framework}{\textsc{CacheTracer}}
\newcommand{\Nflood}{N_{\mathrm{flood}}}
\newcommand{\Nprove}{N_{\mathrm{prove}}}
\newcommand{\Lprove}{L_{\mathrm{probe}}}
\newcommand{\Mcontain}{M_{\mathrm{contain}}}
\newcommand{\Mpartial}{M_{\mathrm{partial}}}
\newcommand{\Mnohit}{M_{\mathrm{nohit}}}
\newcommand{\Eabort}{E_{\mathrm{abort}}}
\newcommand{\Clone}{\textsc{Clone}}
\newcommand{\High}{\textsc{Contained}}
\newcommand{\Partial}{\textsc{Partial}}
\newcommand{\NoRel}{\textsc{NoObserved}}
\newcommand{\Incon}{\textsc{Inconclusive}}
\newcommand{\Fail}{\textsc{Fail}}

\title{Uncovering and Understanding Hidden Dependencies in the LLM API Reseller Ecosystem via Prefix-Cache Side Channels}

\author{
\IEEEauthorblockN{
Zimo Ji\IEEEauthorrefmark{1},
Xin Wei\IEEEauthorrefmark{1},
Congying Xu\IEEEauthorrefmark{1},
Wenyuan Jiang\IEEEauthorrefmark{2},
Xin Yang\IEEEauthorrefmark{3},
Zongjie Li\IEEEauthorrefmark{1},
Yudong Gao\IEEEauthorrefmark{1}, and
Shuai Wang\IEEEauthorrefmark{1}}
\IEEEauthorblockA{\IEEEauthorrefmark{1}Hong Kong University of Science and Technology,
Hong Kong SAR, China}
\IEEEauthorblockA{\IEEEauthorrefmark{2}ETH Z\"urich, Z\"urich, Switzerland}
\IEEEauthorblockA{\IEEEauthorrefmark{3}Zhejiang University, Hangzhou, China}
\IEEEauthorblockA{\{zjiag,zligo,shuaiw\}@cse.ust.hk,
\{xweiba,cxubl,ygaodj\}@connect.ust.hk,
wenyjiang@ethz.ch, lucienyang@zju.edu.cn}
}

\newcommand{\CTRLOBS}{$5{,}059$}
\newcommand{\CTRLREQ}{$5{,}997$}
\newcommand{\CTRLRATE}{$5.9\times10^{-4}$}
\newcommand{\CTRLCI}{$[1.2\times10^{-4},\,1.7\times10^{-3}]$}
\newcommand{\CTRLRETEST}{$1{,}660$}
\newcommand{\CTRLFH}{$3$}

\begin{document}

\maketitle

\begin{abstract}

LLM API resellers have become an important access layer to modern LLM services. However, multi-level resale creates an opaque supply chain: a user's request may traverse undisclosed upstream resellers, each of which can inspect or modify prompts and responses, inducing ecosystem-level confidentiality and integrity risks. 
Existing studies audit individual resellers, but provide little visibility into hidden dependencies across resellers. 
We present \framework{}, the first API-only measurement of such hidden dependencies. Our key insight is to exploit \emph{prefix-cache reuse} as a side channel to measure dependency via cache-reach relations.
\framework{} operationalizes this insight with two primitives: \Flood{} populates fresh cache state through one endpoint, and \Prove{} probes whether another can reuse it while excluding
probe-created hits. 

We then conduct a real-world measurement study with \framework{} on $39$ reseller endpoints, sending $1.1$ million API requests across $636$ endpoint pairs.
Our measurements reveal a deep, concentrated cache-reach structure: $37.1\%$ of measured pairs exhibit shared cache reach, the containment order spans seven layers, and one cache reach is contained within at least $31$ of other nodes. We further find that the recovered structure is model-specific. 
We also evaluate the validity of \framework{} through both real-world consistency checks and controlled experiments. The results show its high reliability and accuracy.
These findings reveal substantial hidden dependencies among seemingly independent API resellers. Such deep and concentrated dependencies can create a large potential blast radius, where a confidentiality or integrity failure along a common upstream path may affect users across multiple downstream resellers.
\end{abstract}

\IEEEpeerreviewmaketitle

\section{Introduction}
\label{sec:intro}

Large language models (LLMs) are increasingly accessed through APIs for question
answering~\cite{chiang2024arena}, coding~\cite{jimenez2024swebench,ji2026coding},
cybersecurity~\cite{zhang2025cybench,ji2025measuring}, and agentic
workflows~\cite{hong2024metagpt,claudecode2026,openclaw2026,ji2026taming}.
Because first-party \emph{API providers} (e.g., OpenAI and Anthropic) may impose
rate limits or regional restrictions~\cite{openai2026countries,openai2026ratelimits},
many users instead obtain access through a third-party \emph{API reseller} (e.g.,
OpenRouter~\cite{menlo2026openrouter}
: an intermediary that relays users' requests to upstream API
providers.
This ecosystem has become substantial.
OpenRouter reports a run rate of nearly $1.5$ quadrillion tokens per year as of
May 2026~\cite{a16z2026stateofai}, while at least $17$
reseller endpoints are used to access LLMs in $187$ published
papers~\cite{zhang2026shadowapi}.
API resellers have become an important access layer for modern LLMs.

\noindent \textbf{Problem.}
However, this exposes users to significant confidentiality and integrity risks.
API resellers occupy a privileged position in the request path: they receive
users' prompts and models' responses in plaintext and can modify either before
forwarding them. Prior studies have found code injection or prompt-secret
exfiltration~\cite{liu2026agent}, model
substitution~\cite{cai2025paying,zhang2026shadowapi,kbf2026}, and price
violations~\cite{gatescope2026}. Furthermore, these risks are amplified by
multi-level resale, in which one reseller obtains API access from another and
resells it downstream~\cite{newapi2026,oneapi2026}. This practice creates an
opaque supply chain: each intermediary on the request path can inspect or alter
prompts and responses, while a compromised upstream reseller can affect downstream
traffic routed through it. Multi-level resale can therefore extend users' trust
boundary beyond the reseller they directly choose and make the risks
difficult to assess.

\noindent\textbf{Research goal.}
A complete view of this opaque trust boundary would identify the supplier edges
behind each reseller. Ordinary users, however, generally observe only black-box
API behavior~\cite{cai2025paying,zhang2026shadowapi}, while resellers' internal
routing remains undisclosed~\cite{gatescope2026}. We therefore do not attempt to
recover supplier edges. Instead, we target \emph{dependency visibility}: we
measure externally observable relations across resellers and analyze
their implications for potential supply-chain exposure. In this paper,
\emph{hidden dependencies} refer to the cache-reach relations defined next, not to
verified supplier edges.

\noindent\textbf{Measurement object.}
During autoregressive inference, serving systems can cache the key--value state of
a prompt prefix and reuse it for a later request with the same
prefix~\cite{zheng2024sglang,gu2025auditing}. Commercial APIs from OpenAI,
Anthropic, and Azure isolate this state across provider-defined scopes and report
cache reuse in response
telemetry~\cite{openai2024promptcache,anthropic2024promptcache,azure2024promptcache}.
We call each isolated prefix-cache scope a \emph{cache domain}. For a fixed model,
an endpoint's \emph{cache reach} is the set of cache domains that its requests can
reach. Two endpoints have \emph{shared cache reach} when their cache reaches
overlap. At a selected measurement resolution, if endpoint $A$'s cache reach is
contained in endpoint $B$'s, we call this relation \emph{cache-reach
containment}. Shared cache reach and cache-reach containment together form the
\emph{cache-reach structure} that we measure.

\begin{figure}[t]
\centering
\includegraphics[width=0.96\columnwidth]{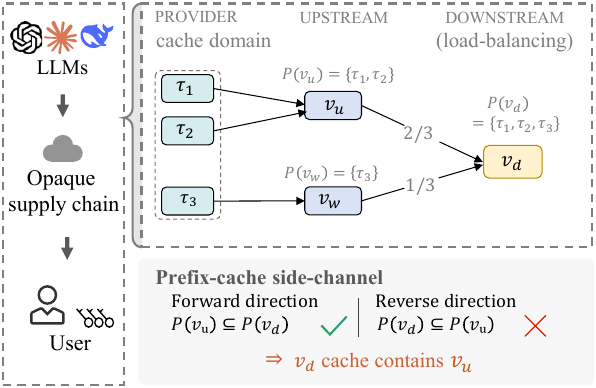}
\caption{Multi-level cache containment and how prefix-cache sharing exposes it. $v_d$, $v_u$ and $v_w$ are resellers; $P(v)$ denotes $v$'s reachable cache domains.}
\label{fig:motivation}
\end{figure}

\noindent\textbf{Measurement method.}
To obtain such evidence, we use prefix-cache reuse as a side channel. The key
\emph{insight} is that prefix-cache reuse makes the cache-reach relations above
externally observable (Section~\ref{sec:insight}). If a fresh prefix cached only
through one endpoint is later reused through another, the two endpoints can reach
a common cache domain, providing evidence of shared cache reach. Multi-level resale can further make this signal
directional. 

We turn the above idea into \framework{}, an API-only measurement method that
probes reseller endpoints pairwise in both directions with two primitives,
\Flood{} and \Prove{} (Section~\ref{sec:design}). For one direction, \Flood{}
repeatedly sends a fresh prompt prefix through endpoint $A$ until the observations
support that the prefix has populated cache domains covering almost all of $A$'s
routing mass (Algorithm~\ref{alg:flood}). \Prove{} then probes endpoint $B$ to
measure whether, and how consistently, $B$ can reuse the cached state populated
through $A$ (Algorithm~\ref{alg:prove}). Because probing can itself populate
cache domains, \Prove{} uses a ladder of previously unqueried prefix extensions so
that newly created state cannot be mistaken for cross-endpoint reuse. Reversing
the roles of $A$ and $B$ classifies the supported cache-reach relation between
them (Algorithm~\ref{alg:det}). 
Repeating this pairwise measurement across all endpoints yields a global
cache-reach structure (Algorithm~\ref{alg:build}). This measurement method
requires only ordinary API access without reseller cooperation or access to
internal reseller systems (Section~\ref{sec:threat}).

The output is one-sided under-approximation (\S\ref{sec:discussion}): every reported
relation is supported by positive cache evidence, while an unreported relation is
unknown rather than absent.

\noindent\textbf{Findings.}
We apply \framework{} to $39$ real-world reseller endpoints, covering $636$
endpoint pairs with $1.1$ million API requests. The results show widespread shared
cache reach and a deep, concentrated cache-reach structure.
\begin{enumerate}[leftmargin=1.2em,itemsep=2pt]
\item $236/636$ ($37.1\%$) of the measured endpoint pairs show shared cache reach.
This result shows that endpoints with different identities can reach common
cache domains. 
\item After endpoints with equivalent cache reach are contracted into $34$ nodes,
the cache-reach containment order is highly concentrated: at least $31$ other
nodes have cache reaches that contain one node's cache reach.
\item The cache-reach containment order spans seven layers. This result is
consistent with multi-level sourcing and trust paths that extend beyond the
reseller selected by the user, although it does not measure commercial hop count.
\item The cache-reach structure is model-specific: among $22$ decisive retests of
cache-reach containment on a second model, only two retain cache-reach
containment. 
\end{enumerate}

These findings show that a confidentiality or integrity failure through a certain endpoint 
may therefore have consequences for more downstream endpoints. Meanwhile, a measurement for one model does not automatically apply to
another, making the hidden dependencies even more complex.

\noindent\textbf{Validation.}
Further validation experiments find only three false-hit reports among $5{,}059$ cold-prompt
observations and show that $14/21$ three-node cache-reach containment chains agree
with the predicted hit probability, both validate our measurement assumptions. In a controlled evaluation,
\framework{} recovers all eight configured
cache-reach structures exactly as well.

In summary, this paper makes the following contributions.
\begin{enumerate}[leftmargin=1.2em,itemsep=2pt]
\item \textbf{Novel problem.} To the best of our knowledge, we are the first to
  systematically study the hidden dependencies created by multi-level resale among
  LLM API resellers. We formulate dependency visibility as an API-only measurement
  problem while distinguishing observable cache-reach relations from unverified
  supplier edges.
\item \textbf{Side-channel insight.} We identify prefix-cache reuse as a side channel for
  measuring shared cache reach and cache-reach containment across 
  reseller endpoints, thereby exposing hidden dependencies in the LLM
  API reseller ecosystem.
\item \textbf{Black-box measurement.} We develop \framework{}, an API-only method to measure pairwise cache-reach relations and
  assemble them into a cache-reach structure. It requires only a paid API key,
  without reseller cooperation or privileged access.
\item \textbf{Empirical findings.} We validate the reliability and effectiveness of
  \framework{} and apply it to $39$ real-world reseller endpoints. The measurements
  reveal widespread shared cache reach and a seven-layer, highly concentrated,
  model-specific containment order, providing evidence of
  ecosystem hidden dependencies exposure. 
\end{enumerate}

These findings can motivate and facilitate further research on supply-chain-aware
security assessment and greater transparency around hidden dependencies in the
LLM API reseller ecosystem. 

\section{Background and Key Observation}
\label{sec:observation}

\subsection{Provider-side cache isolation}
\label{sec:observation:cache}

\begin{figure}[t]
\centering
\includegraphics[width=0.96\columnwidth]{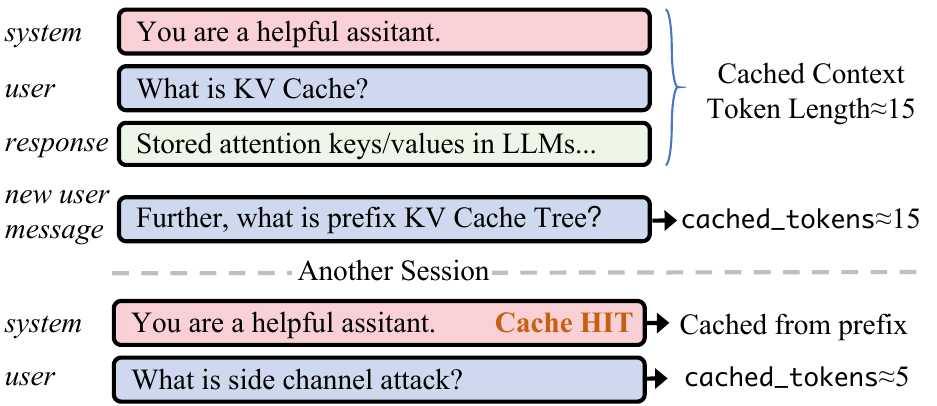}
\caption{Prefix-cache in KV cache tree.}
\label{fig:kvcache}
\end{figure}

Autoregressive inference prefills a prompt into key/value state, then decodes.
Fig.~\ref{fig:kvcache} shows how reuse appears to the customer: a follow-up
request can reuse the full cached conversation prefix, whereas a new session
sharing only the system prompt reuses only that shorter prefix. The API reports
the reused length as \texttt{cached\_tokens}.
Because prefill dominates cost for long prompts, every production stack reuses
KV state across requests sharing a token prefix: PagedAttention in
vLLM~\cite{kwon2023vllm}, RadixAttention in SGLang~\cite{zheng2024sglang}, and
the prompt-cache features of the major commercial
APIs~\cite{openai2024promptcache,anthropic2024promptcache,azure2024promptcache}.

Providers were aware that reuse across customers would be a channel, and closed
it by two means. Each partitions the cache domain by a coarse tenancy identifier: an
organization, project, or workspace; furthermore, they create and store cache in a length unit (chunk, e.g, 256 or 512 tokens)~\cite {openai2024promptcache,anthropic2024promptcache,azure2024promptcache}.
State cached under one tenancy is not reusable from another even when both are
scheduled on the same physical instance. Independent auditing supports this
posture~\cite{gu2025auditing}. Within one provider the property is what it
claims to be: a customer cannot learn anything about another customer's prompts
from the cache.

In our measurement, a successful request is defined as a \Hit\ when its reported cached-token count shows reuse
of the planted prefix, and a \Miss\ otherwise. Transport errors and timeouts
carry no cache-state information and are not counted as either. 
An 
assumption is needed before a hit can be treated as evidence about shared
infrastructure.

\begin{assumptionbox}
\begin{assumption}[Cache Telemetry integrity]
\label{ass:telemetry}
The reported cached-token count reflects genuine reuse of state deposited by
our earlier request.
\end{assumption}
\end{assumptionbox}

This assumption is needed because the customer observes an API field, not the
cache itself.
Economically motivated misreporting is more likely to suppress hits and charge for full input. Such behavior can only remove relations from our results, not create spurious ones, preserving our under-approximation (see \S\ref{sec:discussion}).

\subsection{Reseller-side load balancing}

A reseller is an OpenAI-compatible endpoint that accepts requests from its
customers and reissues them to an upstream service under credentials controlled
by the reseller. Each hop therefore uses two distinct credentials: the customer
calls the reseller with a reseller-issued access key, while the reseller calls
its upstream with an upstream-issued key that the reseller holds. The upstream
sees the reseller's credential rather than the customer's; if the upstream is
itself a reseller, the same credential replacement occurs again. This separation
is explicit in OpenAI-compatible gateways through customer-facing tokens and
provider-facing channels~\cite{oneapi2026}.

A reseller can configure multiple upstream channels rather than a single
credential. New API supports weighted channel selection and automatic retry,
while LiteLLM provides load balancing and fallback across multiple
deployments~\cite{newapi2026,litellmrepo2026}. Consequently, repeated requests
sent through one endpoint may be re-authenticated under
different upstream credentials and reach different provider-side cache
domains.

\noindent \textbf{Reseller Routing model.} For a fixed model, we abstract the reseller's channel selection, retry, and fallback behavior by the induced distribution ($\pi_v$) over provider-side cache domains for endpoint ($v$). The distribution need not be uniform: different domains may carry arbitrarily different fractions of the endpoint's successful requests.

\begin{assumptionbox}
\begin{assumption}[Reseller routing consistency]
\label{ass:routing}
During one measurement window, successful requests to an endpoint are independent samples from a fixed routing distribution ($\pi_v$). 
Moreover, the prompt prefixes of cached prompts do not systematically change this distribution.
\end{assumption}
\end{assumptionbox}

This assumption makes cache observations across repeated requests comparable. Temporal stability allows finite sequences of hits and misses to estimate routing mass, while prefix invariance allows observations made with different prompt prefixes in \framework{}~to characterize the same endpoint-level routing.
Common gateways select eligible upstream channels using model availability, configured weights or priorities~\cite{newapi2026,litellmrepo2026}.
These signals are independent of prompt itself, so both sample the same channel distribution. \S\ref{sec:eval:assumptions} further validates this assumption.

\begin{figure*}[!ht]
  \centering
\includegraphics[width=\textwidth]{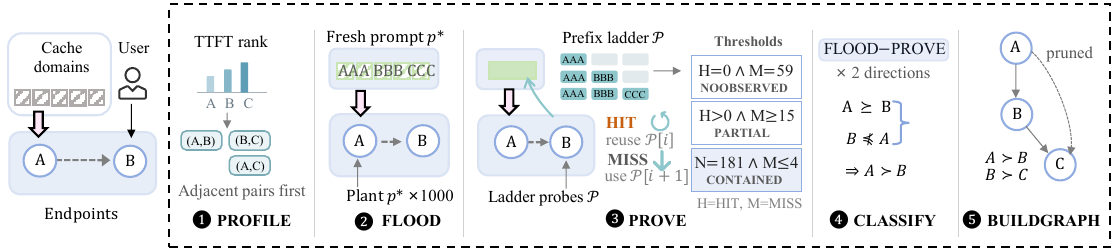}
\caption{Overview of \framework{}. 
}
\label{fig:algorithm}
\end{figure*}

\subsection{How reseller routing exposes the cache channel}
\label{sec:insight}

Fix a prompt and a measurement window. Let $\mathcal{T}$ be the set of cache
domains at which the provider partitions its prefix cache, and let $\Pool(v)$
be the domains in which a request issued at endpoint $v$ can deposit or retrieve
cache state. The prompt and window remain implicit below. Two facts about this
set drive our work.

\begin{enumerate}[leftmargin=*]

\item \textbf{$\Pool$ is monotone along the supply relation.} If $v_d$ forwards
some of its traffic through $v_u$, then every cache domain $v_u$ reaches is also
reachable from $v_d$, through $v_u$: $\Pool(v_u) \subseteq \Pool(v_d)$. The
inclusion is strict exactly when $v_d$ also buys from someone else.

\item \textbf{Load balancing makes the whole set reachable.} A
customer of $v_d$ does not choose which upstream credential carries a request;
the balancer does, and over many identical requests it exercises the whole set.
Every cache domain in $\Pool(v_d)$ carries a lower-bounded share of the traffic.

\end{enumerate}
Together, these facts create a directional channel, illustrated in
Fig.~\ref{fig:algorithm}. 
Take a prompt $p$ minted fresh, never sent anywhere.
Send it repeatedly to the \emph{downstream} $v_d$ until every
cache domain it reaches holds $p$. Because $\Pool(v_u) \subseteq \Pool(v_d)$,
this has planted $p$ in every domain the \emph{upstream} $v_u$ can
reach, even though $p$ was never sent to $v_u$ and $v_u$ has no
relationship with us. Every subsequent probe at $v_u$ is a cache hit. Run the
experiment the other way and it degrades: flooding $v_u$ covers only
$\Pool(v_u)$, so probes at $v_d$ hit only when its balancer selects that subset
of its reach, at a rate below one. Thus, without crossing any provider-side
cache isolation boundary, reseller routing turns cache-reach containment into a
directional signal.

\paragraph{Measurement target.}
Let $\V$ be the endpoints accessible to the measurer. After a fresh prompt
$p$ has been planted through $v_x$, define the \emph{coverage}
$c(v_x,v_y)$ as the probability that a subsequent successful query at $v_y$
hits that state---equivalently, the share of $v_y$'s routing mass covered by the
namespaces reached through $v_x$. At resolution $\delta=0.05$, we write
$v_y\preceq v_x$ when $c(v_x,v_y)\ge1-\delta$, and write $v_y\prec v_x$ when
the reverse containment does not hold. Containment concerns routing mass, not
literal inclusion of zero-probability paths, and does not by itself establish a
commercial supplier. Section~\ref{sec:design} turns this population property
into a finite-sample test. 
\paragraph{Why we choose $\delta=0.05$.}
A tighter resolution requires a longer sequence of cache hits, which increases both the request volume and the measurement duration. Higher request volume can cause OpenAI~\cite{openai2024promptcache} and Azure~\cite{azure2024promptcache} to distribute requests across additional cache domains, while a longer measurement increases the chance that cache state in previously hit domains expires~\cite{openai2026ratelimits,newapi2026,litellmrepo2026}. Both effects introduce transient misses even when nearly all routing mass is covered. In preliminary runs, we observed such misses frequently at $\delta=0.01$,
making $99\%$ coverage difficult to distinguish reliably from full containment
under these production effects. We therefore set $\delta=0.05$ and
operationally define cache containment as coverage of at least $95\%$, treating
the remaining routing mass as negligible.

\subsection{Threat and measurement model}
\label{sec:threat}

The measurer is an ordinary paying customer. It holds low-privilege
credentials for each candidate endpoint and issues chat-completion
requests. It observes response payloads, usage metadata, errors, and end-to-end
timing. It does not control the network, the reseller software, the provider
scheduler, model weights, or any other tenant, and it never sees another
customer's data: prompts are synthetic padding plus a fresh nonce.

\section{\framework{}}
\label{sec:design}

Section~\ref{sec:insight} presents that cache-reach containment induces an observable directional signal. 
Turning this signal into a practical measurement procedure, however, requires addressing two challenges: (i) cache domains cannot be enumerated from outside, and (ii) each probe can change the cache it observes.
To address these challenges, we develop \framework{}, an API-only measurement framework consisting of five primitives that progressively recover pairwise cache-reach relationships and assemble them into a global containment structure.

\subsection{Method overview}

\framework{} takes a set of candidate reseller endpoints as input and turns pairwise
cache observations into a global cache-reach structure through five steps
(Figure~\ref{fig:algorithm}). \Profile\ first collects a lightweight latency
profile for each endpoint and uses it to prioritize the order in which endpoint
pairs are tested. For each selected pair, \Flood\ plants a fresh prompt through
one endpoint until its cache state covers the endpoint's routing mass at the
resolution defined in Section~\ref{sec:insight}. \Prove\ then probes the second
endpoint with a prefix ladder and records how much of its
routing mass can reuse the state planted by \Flood. \Determine\ repeats this
measurement in both directions and combines the two directional outcomes to
classify the pair as equivalent cache reach, strict containment, partial
sharing, no observed sharing, or inconclusive. Finally, \Build\ schedules these
pairwise measurements across all endpoints and assembles their outcomes into
the recovered cache-reach structure.

\subsection{\Profile: prioritizing potentially adjacent pairs}
\label{frame:profile}

\paragraph{Goal.}
\framework{} need to avoid measuring a pair when its containment relation is already
implied by relations discovered earlier. For example, if $A\prec B$ and
$B\prec C$ are discovered before $(A,C)$ is tested, then $A\prec C$ is already
implied and the pair can be skipped. If $(A,C)$ is tested first, however, that
measurement has already incurred its request cost.
\Profile\ therefore provides a lightweight heuristic for prioritizing pairs that
are more likely to expose short-range relations first, allowing \framework{} to prune
more redundant long-range tests later.

\paragraph{Procedure and outcomes.}
\Profile\ works in two stages:

\ding{172} \emph{Latency ranking.}
\Profile\ 
sends one short streaming request with prompt 
\texttt{Reply with exactly OK.} (\LatencyTest\ function in Algorithm~\ref{alg:build}) to each endpoint and records its time to first
token (TTFT). We use TTFT only as a coarse proxy for path length: additional
forwarding and gateway processing can increase response latency, so endpoints
with nearby positions in the TTFT ranking are treated as more plausible
candidates for nearby relations. 
\S~\ref{frame:build}
sorts all successfully profiled
endpoints by TTFT in Line~\ref{line:build-profile} of
Algorithm~\ref{alg:build}.

\ding{173} \emph{Pair ordering.}
Given the ranked endpoint list $E=(e_1,\ldots,e_N)$, we use \textsc{PairsByIncreasingStride} (in Algorithm~\ref{alg:build}) function to construct candidate
pairs by increasing rank distance in 
Algorithm~\ref{alg:build}, \S\ref{frame:build}. It first considers pairs one position apart,
$(e_1,e_2),(e_2,e_3),\ldots$; then pairs two positions apart,
$(e_1,e_3),(e_2,e_4),\ldots$; and continues with increasing stride. 

For
example, for a ranking $(A,B,C,D)$, the candidate order begins with
$(A,B),(B,C),(C,D)$, followed by $(A,C),(B,D)$, and finally $(A,D)$.
Thus, \Profile\ does not decide whether two
endpoints are related; it only changes which pairs are tested earlier.

\subsection{\Flood: covering one endpoint's cache reach}

\noindent \textbf{Goal.}
\Flood\ plants a fresh prompt to certain endpoint without being able to
list the cache domains behind it. 
Under assumption~\ref{ass:routing}, \Flood\ works by continuing to send the same prompt repeatedly until the termination condition is met (Algorithm~\ref{alg:flood}, line~\ref{line:flood-send}).
At
our setting $\delta=0.05$, \Flood\ can stop (line~\ref{line:flood-stop}) when any
uncovered routes together carry less than $5\%$ of the traffic.

\noindent \textbf{\Flood~parameters.}
\Flood~(Algorithm~\ref{alg:flood}) stops after $\Nflood$ 
consecutive 
hits (line~\ref{line:flood-stop}), aborts after $1000$ total attempts or eight consecutive transport errors (line~\ref{line:flood-terminate}),
and then starts two staggered low-rate workers to keep the planted state warm
during probing (line~\ref{line:flood-keepwarm}). We choose $\Nflood$ directly from the target resolution
$\delta=0.05$ and significance level $\alpha=0.05$. Under
Assumption~\ref{ass:routing}, successful requests during a flood sample the
same routing distribution. If at least a $\delta$ fraction of $A$'s routing
mass were still uncovered, i.e., $q_A \ge \delta$, then the probability that
the next request is a hit is at most $1-\delta$, and the probability of
observing $n$ consecutive hits is therefore at most $(1-\delta)^n$. To make
such a hit streak occur with probability at most $\alpha$ under
$q_A\ge\delta$, we choose the smallest integer $n$ satisfying
\[
    (1-\delta)^n \le \alpha,
    \qquad\text{or equivalently}\qquad
    n \ge \frac{\ln \alpha}{\ln(1-\delta)}.
\]
Thus,
\[
    \Nflood
    = \left\lceil \frac{\ln \alpha}{\ln(1-\delta)} \right\rceil
    = \left\lceil \frac{\ln 0.05}{\ln 0.95} \right\rceil
    = 59.
\]
Hence, observing $59$ consecutive hits supports that less than $5\%$ of
$A$'s routing mass remains uncovered; consequently, no individual uncovered
cache domain can account for $5\%$ or more of that routing mass.

\paragraph{Judging cache hit.}
Providers report cached tokens at model-specific cache-chunk granularity, so a
fully cached prompt may still have fewer reported cached tokens than its
total length. We therefore count a response as a hit (line~\ref{line:flood-hit}) when
\[
\texttt{cached\_tokens} \ge |p|-b_B,
\]
where $|p|$ is the length of send prompt $p$, $b_B$ is the cache-chunk length for model $B$, measured during calibration
(\S\ref{sec:eval:sku}).

\paragraph{Keep cache warm.}
$\textsc{StartStaggeredKeepWarm}(A,p,w)$ (line~\ref{line:flood-keepwarm}) starts $w$ low-rate background workers
that repeatedly send the same planted prompt $p$ through $A$ while \Prove\ runs, reducing the
chance that the planted cache state expires. In our setting,
$w=2$ is the calibrated keep-warm concurrency. 
Empirically, we find that this setting can keep the cache warm without 
causing the measurement process to fail frequently due to an excessive 
number of requests.

\begin{algorithm}[H]
\small
\caption{$\Flood(A,p)$: cover $A$'s cache reach with $p$.}
\label{alg:flood}
\begin{algorithmic}[1]
\State $r\gets0$; $a\gets0$; $e\gets0$
\While{$r<\Nflood$} \label{line:flood-stop}
  \If{$a=1000 \lor e=8$} \Return \Fail\ \label{line:flood-terminate} \EndIf
  \State $o\gets\textsc{SendPrompt}(A,p)$; $a\gets a+1$ \label{line:flood-send}
  \If{$o=\textsc{Error}$} $e\gets e+1$; \textbf{continue} \EndIf
  \State $e\gets0$
  \If{$o.\texttt{cached\_tokens}\ge |p|-b_B$} $r\gets r+1$ \label{line:flood-hit}
  \Else{} $r\gets0$ \EndIf
\EndWhile
\State \Return $\textsc{StartStaggeredKeepWarm}(A,p,2)$ \label{line:flood-keepwarm}
\end{algorithmic}
\end{algorithm}

\subsection{\Prove: directional cache-reach test}

\paragraph{Goal and challenge.}
After \Flood\ plants a fresh prompt through endpoint $A$, \Prove\ measures the
directional coverage $c(A,B)$: the probability that a subsequent successful
request through $B$ can reuse the cache state planted through $A$.
We use \Prove\ to denote the complete directional procedure and
\emph{probe} for one cache query issued within it.

\begin{figure}[t]
\centering
\includegraphics[width=0.96\columnwidth]{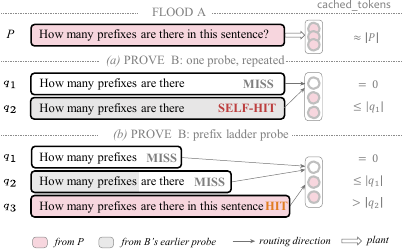}
\caption{Self-hit and prefix ladder of \Prove.}
\label{fig:selfhit}
\end{figure}

A naive implementation cannot repeatedly send the same prompt to $B$.
If a probe misses, that request may itself populate the corresponding cache
domain; a later probe routed to the same domain could then hit state created by
the earlier probe rather than by \Flood. We call this \emph{self-hit} (see Figure~\ref{fig:selfhit} a). Therefore, after each miss, \Prove\  needs to move subsequent probes to
cache state that has never been queried through $B$ itself (Algorithm~\ref{alg:prove}, line~\ref{line:prove-advance}).

\paragraph{Prefix ladder.}
To avoid self-hit, \Prove\ uses a ladder of nested prefixes of prompt $\mathcal{P}$:
\[
    \mathcal{P}[1] \subset \mathcal{P}[2] \subset \cdots
    \subset \mathcal{P}[\Lprove].
\]
For example, as illustrated in Figure~\ref{fig:selfhit}b, let
$q_1 \subset q_2 \subset q_3=P$ be increasingly longer prefixes of prompt $P$.
\Flood\ sends only the full prompt $P$ through $A$.
Because caching $P$ also prepares the KV-cache chunks shared with its shorter
prefixes (see~\S\ref{sec:observation:cache}), \Prove\ can probe these prefixes
through $B$ without planting them through $A$ separately.
Suppose $q_1$ misses. That request may itself cache $q_1$ in the cache domain
to which it is routed. If $q_2$ is later routed to the same domain, it may
therefore reuse up to $|q_1|$ tokens even though that cache was not planted by
\Flood.  We do not count such reuse as a hit: when probing $q_2$, \Prove\ 
requires the returned \texttt{cached\_tokens} to exceed $|q_1|$.
Thus, a cache entry created by the earlier $q_1$ probe cannot by itself make
$q_2$ pass the hit test.  
$q_3$ is therefore a genuine
hit because its cached prefix extends beyond $q_2$.

Generally, in Algorithm~\ref{alg:prove}, \Prove\ starts from the shortest prefix (line~\ref{line:prove-init}). After a hit, it 
safely reuses the
current prefix (line~\ref{line:prove-hit}) because the queried state was already present before that probe.
After a miss, the probe 
has created cache state for that prefix,
so \Prove\ permanently consumes it and advances to the next longer, previously
unqueried prefix (line~\ref{line:prove-advance}). A timeout also consumes the current prefix (line~\ref{line:prove-timeout}) because the request
may have reached the backend and populated the cache even though no response
was observed.

The prefix stride is chosen larger than the provider's cache-chunk length
$b_B$. Hence, cache state created by a miss at $\mathcal{P}[\ell]$ cannot
satisfy the hit threshold for $\mathcal{P}[\ell+1]$.

This construction relies on Assumption~\ref{ass:routing}: changing the prefix
must not systematically change the routing distribution sampled by the probe.
Otherwise, observations from different ladder positions would not characterize
the same endpoint-level routing population.
Section~\ref{sec:eval:consistency} validates this assumption empirically.

\paragraph{Procedure and outcomes.}
Algorithm~\ref{alg:prove} applies the prefix ladder to obtain a bounded
directional measurement. It maintains the number of successful probes $n$,
hits $h$, misses $m$, and the current ladder position $\ell$ (line~\ref{line:prove-init}).

For a response $o$ to prefix $p$, \textsc{IsHit} (line~\ref{line:prove-ishit}) returns true when
\[
    o.\texttt{cached\_tokens}
    \ge \textsc{Tokens}(p)-b_B,
\]
allowing for the final provider cache chunk that may be omitted from the
reported cached-token count. A hit increments $h$ and keeps the current prefix (line~\ref{line:prove-hit});
a miss increments $m$ and advances the ladder (line~\ref{line:prove-advance}). Transport errors do not provide
a cache observation (line~\ref{line:prove-transport}), while a timeout additionally consumes the current prefix
because it may have populated cache state (line~\ref{line:prove-timeout}).

\Prove\ returns one of four directional outcomes. 
\begin{itemize}[topsep=0pt,leftmargin=*]
\item \High\ (line~\ref{line:prove-high}) indicates that the
observed miss count is sufficiently small to support containment at the target
resolution. 
\item \NoRel\ (line~\ref{line:prove-norel}) indicates that no hit was observed before the terminate condition. 
\item \Partial\ (line~\ref{line:prove-partial}) indicates that at least one hit established shared cache
reach but sufficiently many misses reject containment. 
\item All other executions,
including exhaustion of clean prefixes (line~\ref{line:prove-exhaust}) or repeated transport failures (line~\ref{line:prove-abort}), return
\Incon\ (line~\ref{line:prove-incon}).
\end{itemize}

\paragraph{\Prove\ parameters.}
We set the measurement resolution and per-direction significance level to
$\delta=\alpha=0.05$. Under Assumption~\ref{ass:routing}, successful probes
within one \Flood--\Prove\ direction are independent samples from the same
routing distribution, including when \Prove\ advances across different
prefixes in the ladder. 
We choose the following thresholds so that each directional decision has error below~$\alpha$~at~resolution~$\delta$.

\ding{172} \emph{\High\ threshold.}
The \High\ outcome is intended to support that the cache state planted through
$A$ covers $B$ at the target resolution. We allow a small number of misses
rather than require a perfect hit trace. 
For a probe sequence
of $N$ successful requests, the probability of observing at most $m$ misses is
maximized at $\delta$ and is therefore bounded by
\[
\Pr[\operatorname{Bin}(N,\delta)\le m]
\le
\sum_{k=0}^{m}
\binom{N}{k}\delta^k(1-\delta)^{N-k}.
\]
For a given miss allowance $m$, we choose the smallest probe budget $N_m$
for which this probability is at most $\alpha$:
\[
N_m
=
\min \{
N:
\sum_{k=0}^{m}
\binom{N}{k}\delta^k(1-\delta)^{N-k}
\le \alpha
\}.
\]
Allowing $0$, $2$, $4$, $6$, and $8$ misses requires respectively
$59$, $124$, $181$, $234$, and $286$ successful probes. We choose
$\Mcontain=4$ as a practical compromise: it tolerates occasional misses while
keeping the directional probe sequence below $200$ successful requests.
Allowing more misses would substantially increase the probe budget, increasing
both measurement cost and the time over which the state planted by \Flood\
must remain available. This gives
\[
\Nprove=181.
\]
Consequently, if \Prove\ completes $\Nprove=181$ successful probes (line~\ref{line:prove-budget}) with at
most $\Mcontain=4$ misses, it returns \High\ (line~\ref{line:prove-high}).

\ding{173} \emph{\NoRel\ threshold.}
The \NoRel\ outcome addresses the opposite question: how many misses without
any hit are needed to determine that no shared cache reach carrying at least $\delta$ of endpoint $B$'s routing mass?
The probability of
observing $m$ consecutive misses before the first hit is then at most
\[
(1-\delta)^m.
\]
We therefore choose the smallest integer $m$ such that
\[
(1-\delta)^m\le\alpha,
\]
giving
\[
\Mnohit
=
\left\lceil
\frac{\ln\alpha}{\ln(1-\delta)}
\right\rceil
=
\left\lceil
\frac{\ln0.05}{\ln0.95}
\right\rceil
=59.
\]
Thus, if the first $\Mnohit=59$ successful probes are all misses, \Prove\
returns \NoRel\ (line~\ref{line:prove-norel}). This outcome supports that no shared cache reach carrying at
least $\delta$ of $B$'s routing mass was observed. It does not establish that
the cache reaches of $A$ and $B$ are disjoint; sharing below the measurement
resolution may remain unobserved.

\ding{174} \emph{\Partial\ threshold.}
Once at least one hit has been observed, shared cache reach has been witnessed.
The remaining question is whether the miss rate is too large to support
containment at resolution $\delta$. 
Since our containment definition itself allows up to a
$\delta$ fraction of $B$'s routing mass to remain uncovered, observing a small
number of misses is still consistent with containment; we therefore cannot
return \Partial\ as soon as the first miss appears. Instead, we reject
containment only when the number of misses becomes sufficiently large that such
an outcome would be unlikely if containment actually held.
With the probe budget fixed at
$\Nprove=181$, we choose the smallest miss count:
\[
\Mpartial
=
\min\{
m:
\sum_{k=m}^{\Nprove}
\binom{\Nprove}{k}
\delta^k(1-\delta)^{\Nprove-k}
\le\alpha
\}.
\]
Substituting $\Nprove=181$ and $\delta=\alpha=0.05$ gives
\[
\Mpartial=15.
\]
Therefore, after at least one hit has established shared cache reach,
accumulating $\Mpartial=15$ misses is sufficient to reject containment at the
target resolution, and \Prove\ returns \Partial\ (line~\ref{line:prove-partial}). 

\emph{Prefix bounds.}
Each miss may populate the prefix that it probes and therefore consumes one
previously clean ladder position. The longest no-hit decision requires
$\Mnohit=59$ such prefixes, so we set
\[
\Lprove=\Mnohit=59.
\]
A timeout also consumes the current prefix (line~\ref{line:prove-timeout}) because the request may have reached
the backend and populated its cache even though no response was received.
Timeouts do not contribute to the successful-probe count or to the hit and miss
counts (line~\ref{line:prove-count}); if they exhaust the clean prefix ladder, \Prove\ conservatively returns
\Incon\ (line~\ref{line:prove-exhaust}).

Other transport errors likewise carry no cache observation and do not enter
$n$, $h$, or $m$ (line~\ref{line:prove-transport}). We set $\Eabort=8$ consecutive errors (line~\ref{line:prove-abort}) to tolerate short
failure bursts while bounding the time spent on an unavailable endpoint.
The model-specific prefix lengths, prefix stride, and cache-chunk length
$b_B$ are calibrated separately for each measured model
(\S\ref{sec:eval:wild}).

\begin{algorithm}[H]
\small
\caption{$\Prove(B,\mathcal{P})$: probing over prefix ladders.}
\label{alg:prove}
\begin{algorithmic}[1]
\Require clean prefix ladder $\mathcal{P}[1{:}\Lprove]$ and maximum number of
         successful probes $\Nprove$
\Require miss boundaries $\Mcontain,\Mpartial,\Mnohit$ and error cap $\Eabort$
\Function{IsHit}{$o,p$}
  \State \Return $o.\texttt{cached\_tokens}\ge\textsc{Tokens}(p)-b_B$ \label{line:prove-ishit}
\EndFunction
\State $(n,h,m)\gets(0,0,0)$; $\ell\gets1$; $e\gets0$ \label{line:prove-init}
\While{$n<\Nprove$} \label{line:prove-budget}
  \If{$\ell>\Lprove$} \Return $(\Incon,(n,h,m))$ \label{line:prove-exhaust} \EndIf
  \If{$e=\Eabort$} \Return $(\Incon,(n,h,m))$ \label{line:prove-abort} \EndIf
  \State $o\gets\textsc{SendPrompt}(B,\mathcal{P}[\ell])$
  \If{$o\in\{\textsc{Error},\textsc{Timeout}\}$} \label{line:prove-transport}
    \State $e\gets e+1$
    \If{$o=\textsc{Timeout}$} $\ell\gets\ell+1$ \label{line:prove-timeout} \EndIf
    \State \textbf{continue}
  \EndIf
  \State $e\gets0$; $n\gets n+1$ \label{line:prove-count}
  \State $S\gets\Call{IsHit}{o,\mathcal{P}[\ell]}$
  \If{$S$} $h\gets h+1$ \label{line:prove-hit}
  \Else{} $m\gets m+1$; $\ell\gets\ell+1$ \label{line:prove-advance} \EndIf
  \If{$h=0\land m=\Mnohit$} \Return $(\NoRel,(n,h,m))$ \label{line:prove-norel} \EndIf
  \If{$h>0\land m\ge\Mpartial$}
    \Return $(\Partial,(n,h,m))$ \label{line:prove-partial} \EndIf
\EndWhile
\If{$m\le\Mcontain$} \Return $(\High,(n,h,m))$ \label{line:prove-high}
\Else{} \Return $(\Incon,(n,h,m))$ \label{line:prove-incon} \EndIf
\end{algorithmic}
\end{algorithm}

\vspace{-20pt}
\subsection{\Determine: bidirectional pair classification}
\label{sec:classify}

\begin{figure}[t]
\centering
\includegraphics[width=0.9\columnwidth]{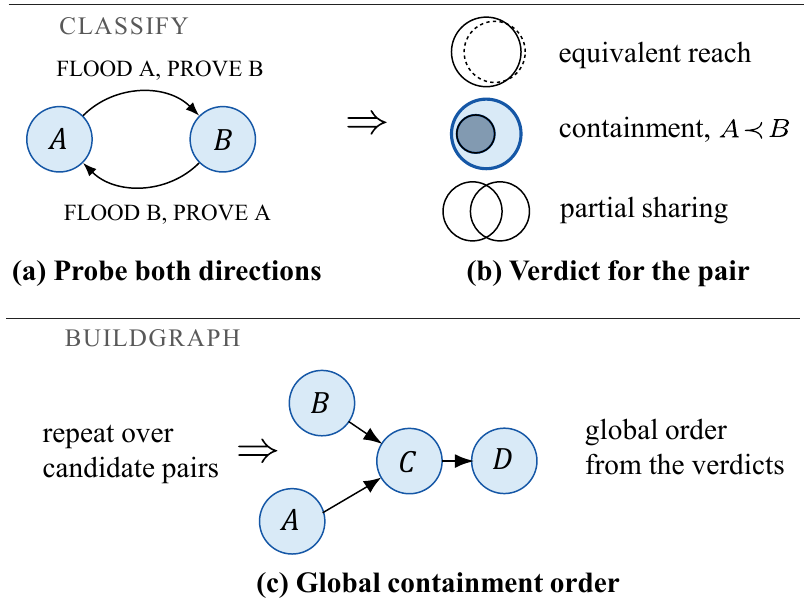}
\caption{From pairwise probes to a global containment order.}
\label{fig:pairwise}
\end{figure}

One flood--probe direction can support $B\preceq A$, but cannot tell whether
the reverse containment also holds. \Determine~(Algorithm~\ref{alg:det}, Figure~\ref{fig:pairwise} a,b)~therefore measures both
directions with independent fresh prompts (lines~\ref{line:det-ladder}--\ref{line:det-prove}, run in both orders at lines~\ref{line:det-dirab}--\ref{line:det-dirba}):
\begin{itemize}[topsep=0pt]
    \item Two \High\ outcomes form a
cache-reach equivalence class (\Clone, line~\ref{line:det-clone}).
    \item One \High\ outcome gives strict
containment when the reverse direction rejects containment (lines~\ref{line:det-containab}--\ref{line:det-containba}); 
    \item A \Partial\ outcome
without containment in either direction records non-directional sharing (line~\ref{line:det-partial}).
    \item Two
\NoRel\ outcomes record no sharing observed at the chosen resolution (line~\ref{line:det-norel}). 
    \item Any
\Incon\ direction leaves the pair unresolved (lines~\ref{line:det-floodfail}, \ref{line:det-incon}). 
\end{itemize}
We write
$X\rightsquigarrow Y$ for the direction that floods $X$ and then probes $Y$.

\begin{algorithm}[H]
\small
\caption{$\Determine(A,B)$: bidirectional  classification.}
\label{alg:det}
\begin{algorithmic}[1]
\Procedure{RunDirection}{$X,Y$}
  \State $\mathcal{P}\gets\textsc{FreshPrefixLadder}_Y(\Lprove)$ \label{line:det-ladder}
  \State $K_X\gets\Flood(X,\mathcal{P}[\Lprove])$
  \If{$K_X=\Fail$} \Return $(\Incon,(0,0,0))$ \label{line:det-floodfail} \EndIf
  \State $(r,\sigma)\gets\Prove(Y,\mathcal{P})$; $\textsc{Stop}(K_X)$ \label{line:det-prove}
  \State \Return $(r,\sigma)$
\EndProcedure
\State $(r_{A\rightsquigarrow B},\sigma_{A\rightsquigarrow B})
       \gets\Call{RunDirection}{A,B}$ \label{line:det-dirab}
\State $(r_{B\rightsquigarrow A},\sigma_{B\rightsquigarrow A})
       \gets\Call{RunDirection}{B,A}$ \label{line:det-dirba}
\If{$r_{A\rightsquigarrow B}=\Incon\lor
      r_{B\rightsquigarrow A}=\Incon$} \Return \Incon \label{line:det-incon} \EndIf
\If{$r_{A\rightsquigarrow B}=\High\land
      r_{B\rightsquigarrow A}=\High$} \Return \Clone \label{line:det-clone}
\ElsIf{$r_{A\rightsquigarrow B}=\High$} \Return $(B\prec A)$ \label{line:det-containab}
\ElsIf{$r_{B\rightsquigarrow A}=\High$} \Return $(A\prec B)$ \label{line:det-containba}
\ElsIf{$r_{A\rightsquigarrow B}=\Partial\lor
        r_{B\rightsquigarrow A}=\Partial$} \Return \Partial \label{line:det-partial}
\Else{} \Return \NoRel \label{line:det-norel} \EndIf
\end{algorithmic}
\end{algorithm}

\paragraph{Containment vs. supply.}
\label{rem:supply}
Monotonicity (\S\ref{sec:insight}) gives supply $\Rightarrow$ containment: if
$v_x$ forwards part of its traffic through $v_y$ then
$\Pool(v_y)\subseteq\Pool(v_x)$. The converse is not available from cache
observations. A reported $v_y \prec v_x$ is equally consistent with
\begin{enumerate}[leftmargin=1.4em,itemsep=1pt,topsep=2pt]
\item $v_y$ being a supplier of $v_x$; and
\item $A$, $B$, and $C$ being upstream providers of $v_x$, while $B$ and $C$
      are upstream providers of $v_y$. In this case,
      $\Pool(v_y)\subsetneq\Pool(v_x)$, so the measurement reports $E\prec D$ even
      though neither $v_x$ nor $v_y$ supplies the other.
\end{enumerate}
Cache-reach containment is evidence that two endpoints may 
connect through the same supply chain, but it does not establish a direct
supplier relationship between them.

\subsection{\Build: pruning and global assembly}
\label{sec:build}
\label{frame:build}

\paragraph{Goal.}
Pairwise measurements produced by \Determine\ must be combined into a global
cache-reach structure. Measuring every candidate pair directly is unnecessary:
some containment relations become implied by relations recovered earlier.
\Build\ therefore processes candidate pairs in the order prioritized by
\Profile\, prunes implied pairs during measurements, and assembles the
remaining pairwise outcomes into the recovered structure (Algorithm~\ref{alg:build}, Figure~\ref{fig:pairwise} c).

\begin{algorithm}[H] 
\small 
\caption{$\Build(\mathcal{L})$: build cache reach structure.
} 
\label{alg:build} 
\begin{algorithmic}[1] 
\State $E\gets\textsc{Sort}(\{e\in\mathcal{L}:\LatencyTest(e)\ne\Fail\})$ by TTFT descending 
\label{line:build-profile} 
\State $R\gets\mathbf{0}$; $\mathcal{K}\gets$ singletons; $\mathcal{P}\gets\emptyset$; $U\gets\emptyset$ \State $Q\gets\textsc{PairsByIncreasingStride}(E)$ 
\label{line:build-pairs} 
\While{$Q\ne\emptyset$} 
\State $(i,j)\gets\textsc{Pop}(Q)$ 
\label{line:build-pop} 
\If{$\textsc{HasPath}(R,i,j)\lor\textsc{HasPath}(R,j,i)$} 
\State \textbf{continue} 
\EndIf 
\label{line:build-skip} 
\State $r\gets\Determine(E[i],E[j])$ 
\label{line:build-determine} 
\If{$r=(E[i]\prec E[j])$} 
\State $R[i][j]\gets1$ \label{line:build-containij} \ElsIf{$r=(E[j]\prec E[i])$} 
\State $R[j][i]\gets1$ \label{line:build-containji} \ElsIf{$r=\Clone$} 
\State $\mathcal{K}.
\textsc{Merge}(i,j)$ \label{line:build-clone} \ElsIf{$r=\Partial$} 
\State $\mathcal{P}\gets\mathcal{P}\cup\{(i,j)\}$ 
\label{line:build-partial} 
\Else 
\State $U\gets U\cup\{(i,j,r)\}$ 
\label{line:build-unresolved} 
\EndIf 
\State delete from $Q$ every pair whose relation is now implied by $R$ 
\label{line:build-prune} 
\EndWhile 
\State $(R,\mathcal{S})\gets\textsc{ShadowInsert}(R,\mathcal{P})$ \label{line:build-shadow} 
\State \Return $(R,\mathcal{K},\mathcal{S},U)$ 
\end{algorithmic} 
\end{algorithm}

\paragraph{Procedure and outcomes.}
Algorithm~\ref{alg:build} first obtains the TTFT-ranked endpoint list $E$ (line~\ref{line:build-profile}) and
the ordered candidate-pair list $Q$ (line~\ref{line:build-pairs}) using the profiling and pair-ordering
procedure described in \S\ref{frame:profile}. It then processes $Q$ from the
front (line~\ref{line:build-pop}). For each pair $(i,j)$ whose relation is not already implied by the
current containment graph $R$ (line~\ref{line:build-skip}), \Build\ invokes
$\Determine(E[i],E[j])$ (line~\ref{line:build-determine}) and records the returned outcome. A strict-containment
verdict adds the corresponding directed relation to $R$ (lines~\ref{line:build-containij}--\ref{line:build-containji}); a \Clone\ verdict
merges the two endpoints in the clone partition $\mathcal{K}$ (line~\ref{line:build-clone}); a \Partial\
verdict is stored in $\mathcal{P}$ for later structural interpretation (line~\ref{line:build-partial}); and
other unresolved outcomes are recorded in $U$ (line~\ref{line:build-unresolved}). After all candidate pairs have
either been measured or pruned, \Build\ calls
\textsc{ShadowInsert} to explain remaining non-directional sharing and returns
$(R,\mathcal{K},\mathcal{S},U)$ (line~\ref{line:build-shadow}).

\paragraph{Implication pruning.}
Whenever a newly recovered containment relation is added to $R$, \Build\ removes
from $Q$ any remaining pair whose relation is already implied by a path in
$R$ (line~\ref{line:build-prune}). For example, once $A\prec B$ and $B\prec C$ have been recovered, the pair
$(A,C)$ no longer requires a direct measurement. This is why the ordering
produced by \Profile\ matters: prioritizing plausibly short-range relations
increases the chance that longer-range pairs become implied before they are
measured. Pruning reduces both request cost and unnecessary cache perturbation;
it does not introduce a new measurement verdict, since every pruned relation is
derived only from relations already stored in $R$.

\paragraph{Latent witnesses for non-directional sharing.}
A \Partial\ verdict indicates shared cache reach without containment in either
direction. Such sharing may already be explained by a common upper endpoint in
the recovered containment structure; otherwise, it may reflect cache reach
through an endpoint that is not among the measured endpoints. To represent the
latter case, \textsc{ShadowInsert} (line~\ref{line:build-shadow}) constructs an auxiliary graph over
unexplained \Partial\ pairs, enumerates its maximal cliques, and introduces one
latent witness for each clique. Cliques may overlap, so one visible endpoint
may participate in multiple explanations, but visible endpoints are never
merged. These witnesses are explanatory objects rather than identified
companies or resellers; Appendix~\ref{app:shadow} establishes their
non-identifiability, and \S\ref{sec:eval} reports them separately from observed
endpoints and containment relations.

Appendix~\ref{app:measurement-budget} analyzes the foreground request complexity of \Build, while \S\ref{sec:eval} reports the realized measurement cost in our field study.

\paragraph{Containment order vs. adjacency}
\label{rem:reach}
A relation $A\prec C$ states only that $A$'s cache reach is contained in
$C$'s; it does not imply that no intermediate endpoint exists between them.
For example, $A\prec B\prec C$ also implies $A\prec C$. Accordingly, once
\Build\ has recovered $A\prec B$ and $B\prec C$, it can skip directly
measuring $(A,C)$ (line~\ref{line:build-skip}).

\section{Measurement \& Evaluation}
\label{sec:eval}

In the evaluation, we aim to answer the following research questions (RQs).

\begin{enumerate}[leftmargin=3em,itemsep=2pt]
\item[\textbf{RQ1}] \textbf{Real-world Measurement.} What cache-containment topology appears in the real-world reseller ecosystem? How does it vary across models, and what security concerns
      does it reveal?
\item[\textbf{RQ2}] \textbf{Assumption Validation.} Are the assumptions for cache observation (\S\ref{sec:design}) consistent with the systems we measured?
\item[\textbf{RQ3}] \textbf{Controlled Evaluation.} Given known ground truth, how accurately does \framework{} recover cache sharing and containment, and what can that structure
      reveal about possible supply relations?
\end{enumerate}

Our primary goal is to characterize hidden dependencies among real-world
resellers (RQ1). 
To assess the validity of the underlying measurement methodology, we examine whether the assumptions underlying \framework{} are consistent with the measured systems 
(RQ2), and evaluate the accuracy of \framework{} against known ground truth in a controlled environment (RQ3).

\subsection{Experiment Setup}
\subsubsection{Prompt construction}
Each prompt consists of the fixed text \texttt{Replay exactly OK} followed by a
nonce string.
Because tokenization varies with prompt content, we use the input-token count
reported in the response's \texttt{usage} field to calibrate prompts to the
target lengths required by our measurement.

\subsubsection{Reseller selection}
\label{sec:eval:selection}
RQ1 and RQ2 use the same set of publicly reachable resellers. 
We identify candidate reseller endpoints from three sources: a FOFA scan for publicly reachable deployments of New
API~\cite{fofa2026search,newapi2026}, a community-maintained relay
directory~\cite{relaydirectory2026}, and TokenAPI Scan's analysis of relay
domains embedded in Claude Code~\cite{tokenscan2026}. 
After deduplication, we
manually registered an account at each candidate, confirmed that a test payment
was credited, and verified that the endpoint returned a usable
\texttt{cached\_tokens} field. This screening retained $39$ endpoints for the
in-the-wild measurements.
Appendix~\ref{app:endpoints} records the selected resellers, and our findings
apply only to this segment.

All requests are ordinary OpenAI-compatible calls from one workstation, with no provider
cooperation and no privileged access to any reseller or its telemetry.

\begin{figure*}[t]
\centering
\includegraphics[width=0.94\textwidth]{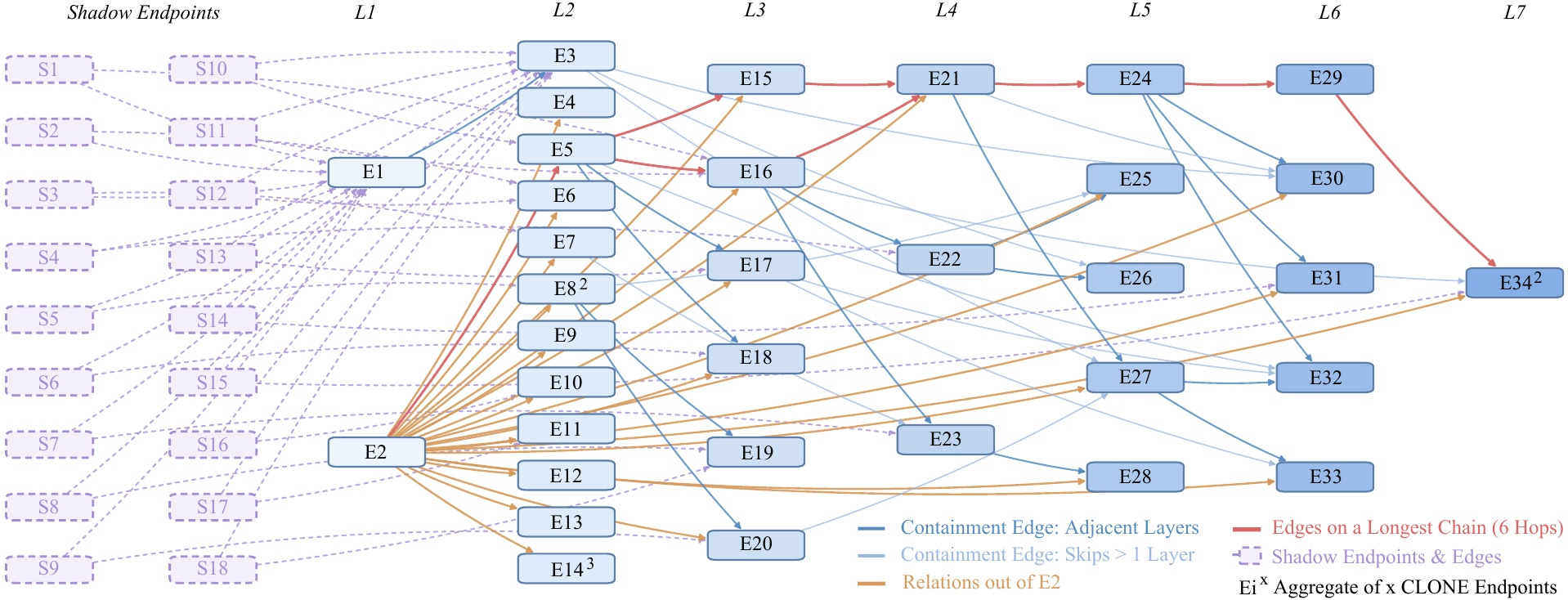}
\caption{Recovered cache-containment order over $39$ 
real-world reseller endpoints.
}
\label{fig:field}
\end{figure*}

\subsection{RQ1: Real-world Measurement}
\label{sec:eval:wild}

\subsubsection{Measurement setup}
We apply \framework{} to these $39$ selected resellers, using \texttt{gpt-5-nano-2025-08-07} as the measured model.
We choose this model rather than a flagship offering for two reasons.
\begin{itemize}[leftmargin=*]
    \item Undisclosed model substitution has been observed among
    resellers~\cite{zhang2026shadowapi}; we therefore use a low-cost model to
    reduce the risk that different endpoints serve different underlying models
    with isolated cache namespaces.
    \item Its low request cost and latency reduce measurement expense and
    shorten the campaign, limiting the effect of routing changes during the
    measurement window.
\end{itemize}

For each measurement, we generate a fresh $16{,}384$-token prompt body and construct a prefix ladder starting at $1{,}536$ tokens with a $256$-token stride. The initial length exceeds the empirically measured $1{,}024$-token cache-admission threshold, while the stride exceeds the measured $64$-token
cache-chunk length. 
Because both parameters are specific to \texttt{gpt-5-nano-2025-08-07}, we recalibrate them before measuring a different model.

The $39$ endpoints yield $741$ unordered candidate pairs. \framework{} measures $636$ pairs over $7.2$ hours and 
prunes $105$ during the measurement.

\begin{table}[t]
\centering
\caption{Results of cache-reach measurements for $636$ real-world reseller endpoint pairs.}
\label{tab:wild-outcomes}
\small
\begin{tabular}{@{}lrr@{}}
\toprule
Measurement outcome & Pair count & Pair share \\
\midrule
Sharing cache reach       & \textbf{236} & \textbf{37.1\%} \\
\quad Non-directional (\Partial)  & 168 & \\
\quad Directed containment        &  62 & \\
\quad Equivalent reach (\Clone)   &   6 & \\
No observed relation at 5\% resolution & 253 & 39.8\% \\
Inconclusive                      & 147 & 23.1\% \\
\midrule
total                             & 636 & 100\% \\
\bottomrule
\end{tabular}
\end{table}

\subsubsection{How common is shared cache reach?}
\label{sec:eval:f2}

Table~\ref{tab:wild-outcomes} shows that $236/636$ measured pairs ($37.1\%$)
produce at least one authentic cross-endpoint hit and a decisive coverage outcome. 
Under the assumptions examined in RQ2, each such
hit provides direct evidence that the two reseller endpoints can reach at least
one common cache domain.

\begin{findingbox}
\noindent\textbf{Finding 1.}
We observe shared cache reach in $236/636$ measured reseller pairs ($37.1\%$), revealing hidden dependencies within the measured segment of the reseller ecosystem. 
\end{findingbox}

\subsubsection{Depth and breadth of cache containment}
\label{sec:eval:graph}

Contracting the six \Clone\ verdicts merges nine endpoints into four equivalence classes, leaving $34$ nodes. 
The $62$ endpoint-level containment verdicts reduce to $58$ distinct relations after contraction.
Fig.~\ref{fig:field} visualizes the resulting relation graph, which is acyclic and spans seven layers.
The graph's longest chain comprises six containment relations, and its transitive closure contains $115$ ordered node pairs.

The seven-node chain provides a direct lower bound on cache-containment depth.
If each strict containment relation corresponded to a distinct resale hop, the
longest chain would be consistent with as many as six resale steps. The topology therefore indicates possible supply-chain depth without measuring commercial hop count.

\begin{findingbox}
\noindent\textbf{Finding 2.}
The observed cache-containment order is at least seven layers deep. This
is direct evidence of nested cache reach and indirect evidence that the reseller
ecosystem may contain multi-level supply paths.
\end{findingbox}

The recovered topology is also highly concentrated.
The cache reach of $E2$ is directly contained in those of $24$ of the other $33$ nodes. Accounting for indirect containment paths, bringing the total to $31$.
The next-largest upward closure also contains $18$ nodes.
Because pairs with no observed relation or inconclusive outcomes contribute no
containment edges, this count is conservative. 
$31/33$ is a lower bound on the number of measured endpoints that can reach the cache domain represented by $E2$. 

\begin{table}[t]
\centering
\caption{Directional evidence for $E2$.}
\label{tab:reverse}
\small
\setlength{\tabcolsep}{4pt}
\begin{tabular*}{\columnwidth}{@{\extracolsep{\fill}}llrr@{}}
\toprule
Flood source & Probe endpoint & Hits/Misses & Num. of pairs \\
\midrule
Other endpoint & $E2$            & $178/3$ & 24 \\
Other endpoint & $E2$            & $177/4$ &  2 \\
$E2$           & Other endpoint  & $0/59$  & 26 \\
\bottomrule
\end{tabular*}
\end{table}

Table~\ref{tab:reverse} provides the directional measurement evidence for the $26$ directly observed endpoint-level containment relations involving $E2$.
When the counterpart endpoint is flooded and $E2$ is probed, all $26$ tests
satisfy the containment criterion. In the reverse direction, however, flooding
$E2$ and probing the counterpart yields zero hits across all $59$ probes for each pair.
Thus $E2$'s cache domain is reachable through $31$
endpoints but carries only a small share of traffic for most of them. This
pattern is consistent both with a narrow supplier and with a co-customer
exposing one tenancy also used by many other resellers.

\begin{findingbox}
\noindent\textbf{Finding 3.}
Cache reach is highly concentrated.
A certain cache domain is contained in the cache reach of at least $31$ of
the other $33$ clone-contracted nodes. 
This establishes a conservative lower bound on cache-reach concentration, indicating a potentially concentrated hidden dependency.
\end{findingbox}

\subsubsection{Security implications}

Cache containment identifies where many nominally independent endpoints can
send traffic through the same cache path. Any operator on that path
receives the context and participates in serving the completion; a
malicious or compromised operator could therefore read prompts or alter returned
content, including tool calls. 

This exposure is particularly consequential for tool-using LLM agents, whose
requests may contain retrieved records, persistent memory, inter-agent
messages, credentials, and tool arguments. Prior work has demonstrated
substantial leakage from agent memory and inter-agent communication, reporting
up to $50.7\%$ leakage of sensitive information despite explicit privacy
instructions and $68.8\%$ leakage in inter-agent messages
~\cite{wang2025memory,juneja2025magpie,elyagoubi2026agentleak}. 
The integrity risk is already concrete: a measurement of $428$ third-party
routers found nine injecting malicious code into agent responses, $17$ using
researcher-controlled AWS canary credentials, and one using a
researcher-controlled private key to drain ETH~\cite{liu2026agent}.

In a resale chain, a malicious or compromised upstream can apply the same
read-and-rewrite capability to any downstream traffic that traverses it. The
depth and concentration we recover therefore imply a potential blast radius
larger than any one public endpoint.

\begin{findingbox}
\noindent\textbf{Finding 4.}
The recovered topology reveals correlated security exposure across nominally
independent resellers. Consequently, a malicious or compromised shared upstream could expose
sensitive context or corrupt tool-mediated actions across multiple downstream
endpoints,
which turns an upstream compromise from an endpoint-local risk into a cross-reseller security concern.
\end{findingbox}

\subsubsection{Model dependence}
\label{sec:eval:sku}
We next examine whether the recovered cache-containment topology is model-specific. 
After recalibrating the cache-admission threshold and cache-chunk size for \texttt{gpt-5.1}, we repeat the measurement on the $44$ endpoint pairs that had a reported containment relation on \texttt{gpt-5-nano-2025-08-07}.

The results show that, the recovered cache-containment topology is model-specific.  
Of these $44$ reruns, $22$ produce a decisive verdict. Only $2$ reproduce the original containment relation; $12$ still
exhibit shared cache reach but no longer satisfy the containment criterion; and
$8$ show no detectable sharing at the $5\%$ resolution. The remaining $22$
reruns are inconclusive and therefore provide no evidence either for or against
cross-model persistence. Fig.~\ref{fig:cross-model} in Appendix~\ref{app:gpt51} shows the two containment
relations reproduced across both models.

\begin{findingbox}
\noindent\textbf{Finding 5.}
Cache-containment topology is model-dependent. 
Only $2$ of the $22$ decisive reruns reproduce the corresponding containment relation observed for the original model. 
The recovered topology should therefore be indexed by model rather than interpreted as a model-agnostic graph.
\end{findingbox}

\paragraph{A possible explanation.}
Industry reports suggest that resellers source different models through
different channels. For flagship models, some pool consumer subscriptions
behind reverse proxies that turn logged-in web or client sessions into an API;
Antigravity similarly translates Anthropic requests to Google's Cloud Code API
and rotates Google accounts as quotas
bind~\cite{yicai2026relay,antigravityproxy2026}. For low-cost API
models, smaller resellers may instead buy token quota from larger
relays~\cite{xiao2026token}. These independently assembled upstream pools could
produce different cache domains.

\subsubsection{Measurement cost}
\label{sec:eval:cost}

Finally, we examine the practical cost of running \framework{} over the
$39$ real-world reseller endpoints. Measuring all candidate pairs requires a
large number of repeated probes, because each pair must be tested in both
directions and each directional test uses repeated observations to determine
whether cache sharing is present.

In total, the real-world measurement sends approximately
$1.1\times10^{6}$ requests over $7.2$ hours. 
This corresponds to about $42$ requests per second on average and approximately
$1{,}730$ requests for each directly measured pair. 
These requests include the measurements in both directions, the two $181$-observation probe sequences used
for each pair, and the additional requests used to keep relevant cache entries
active during measurement.

The implication-pruning mechanism further reduces the measurement cost. Among
the $741$ candidate endpoint pairs, $105$ pairs ($14.2\%$) do not need to be
measured directly because their containment relations can already be inferred
from previously recovered relations.

\subsection{RQ2: Assumption Validation}
\label{sec:eval:assumptions}

\framework{} identifies cache sharing and containment based on the cache telemetry and
hit patterns observed across repeated requests. 
In this RQ, we verify these assumptions on the same $39$ reseller endpoints used in RQ1.

\subsubsection{Telemetry integrity}

Assumption~\ref{ass:telemetry} requires the
\texttt{cached\_tokens} value reported by a reseller to reflect an actual cache
hit. 
We check this by sending
cold prompts to all $39$ endpoints. Every request uses a fresh nonce and a
prompt never previously sent through any endpoint. A reported full-prefix hit
therefore cannot be attributed to cache planted by our requests.

We send \CTRLREQ\ distinct cold-prompt requests and obtain \CTRLOBS\ usable
observations. The results show that three endpoints do not return enough usable telemetry for this
check. Among the remaining endpoints, we observe \CTRLFH\ false hits, corresponding to a rate of \CTRLRATE\ (95\% CI: \CTRLCI).
Repeating the check with refreshed prompts for
\CTRLRETEST\ more observations reproduces one event at one endpoint and none at
the other, leaving endpoint-level rates of $1/603$ and $2/1270$. We treat these
two endpoints as exceptions and exclude their \Partial\ verdicts from aggregate
sharing results.

\begin{findingbox}
\noindent\textbf{Finding 6.}
\emph{Cache telemetry is reliable for nearly all measured endpoints.}
We exclude the two exceptions' partial-sharing verdicts from aggregate results.
\end{findingbox}

\subsubsection{Stable routing}
\label{sec:eval:consistency}

Assumption~\ref{ass:routing} concerns how repeated requests
sample the underlying routing paths. In our black-box setting, we cannot observe
routing decisions directly.
To check its observable consequence,
we run \Build\ on the $39$ selected resellers and examine each recovered chain
$A\prec B\prec C$. For each chain, we independently measure the probability of
a hit when probing $B$ after flooding $A$, probing $C$ after flooding $B$, and
probing $C$ after flooding $A$. If routes from $C$ to $A$ pass through $B$, the
routing model predicts
\[
\begin{aligned}
&\Pr[\text{probe at }C\text{ hits}\mid\text{flood }A] \\
&\quad=\Pr[\text{probe at }B\text{ hits}\mid\text{flood }A] \\
&\qquad\times\Pr[\text{probe at }C\text{ hits}\mid\text{flood }B].
\end{aligned}
\]
The intuition is that, the fraction of $C$'s requests that reach $A$ should equal the fraction
that reach $B$, multiplied by the fraction of $B$'s requests that reach $A$. Each probability is
measured with a separate flood and a fresh random prompt.

\begin{figure}[t]
\centering
\includegraphics[width=0.8\columnwidth]{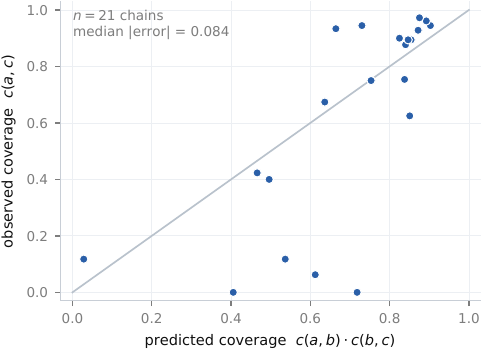}
\caption{Predicted and observed hit probabilities for $21$ 
$A\prec B\prec C$ chains.}
\label{fig:consistency}
\end{figure}

The recovered topology contains $21$ such available chains. Fig.~\ref{fig:consistency} compares the predicted and observed
probability that a probe at $C$ hits after flooding $A$.
The median absolute deviation is $0.084$, with $14/21$ chains agreeing within
$0.10$; randomly permuting the observed long-range values raises the median
deviation to $0.449$. Whole-prompt cache domain pinning would instead make the
three randomized floods sample unrelated routing populations and would not
produce this agreement. The seven larger deviations show that the assumptions
are approximations, not invariants: temporary route changes, quotas, or
alternative paths can still make relations unobservable or make the measured
probabilities inconsistent.

\begin{findingbox}
\noindent\textbf{Finding 7.}
Independently measured hit
probabilities are broadly consistent with the multiplicative relationship
predicted by our routing model. This provides empirical support for stable
routing and prefix-invariant selection as reasonable approximations for the
measured systems.
\end{findingbox}

\begin{figure*}[!t]
  \centering
  \includegraphics[width=0.95\textwidth]{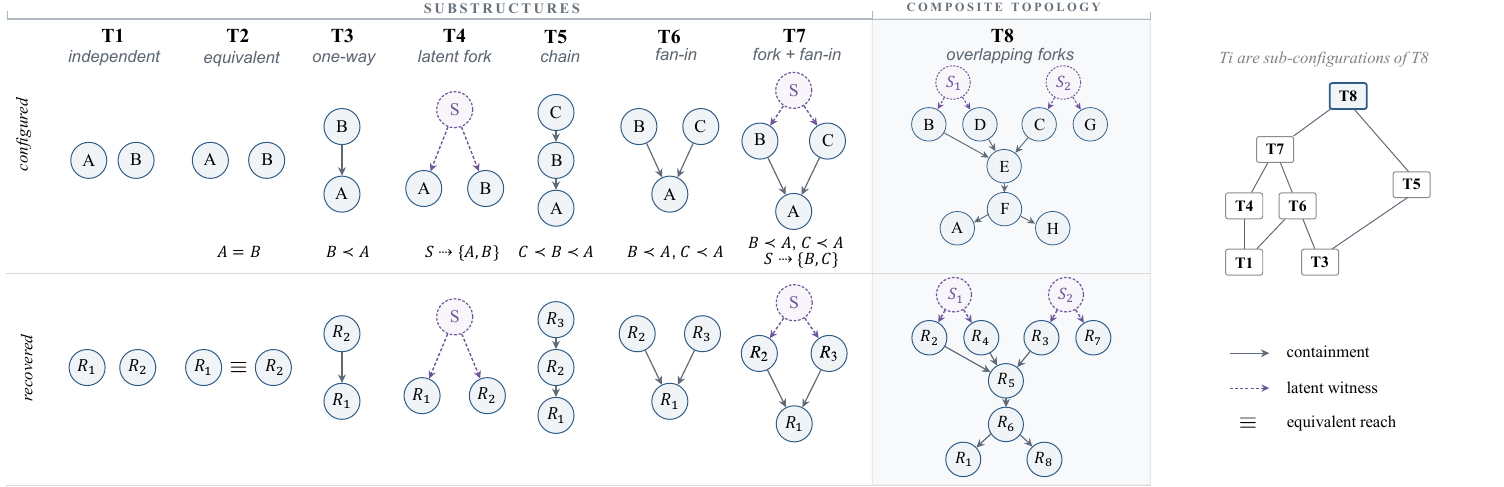}
  \caption{Configured (top) and recovered (bottom) structures for T1--T8. 
}
  \label{fig:suite}
\end{figure*}

\subsection{RQ3: Controlled Evaluation}
\label{sec:eval:controlled}

Real-world reseller endpoints provide no observable ground truth about their
underlying cache-sharing or supply relationships. To evaluate the accuracy of \framework{}, we therefore prepare a controlled environment where the complete topology is known
by construction. 

Specifically, we configure eight topologies using
independently isolated cache domains under our own credentials. Relays
forward requests through these credentials, allowing us to determine exactly
which cache domains each relay can reach and which upstreams are
shared.
For this controlled environment, we set the cache-chunk length to
$32$ tokens, the prompt length to $3{,}481$ tokens, the prefix stride to
$32$ tokens, and the keep-warm concurrency to $4$. The controlled and
real-world evaluations use the same relation classifier and topology-assembly
logic.

\noindent\textbf{Experimental topologies.}
We design the eight topologies to progressively evaluate \framework{} (Fig.~\ref{fig:suite}).
\begin{itemize}[topsep=0pt]\item T1--T4 cover the four pairwise outcomes returned by \Determine: no observed sharing, equivalent cache reach, strict containment, and sharing without containment. 
    \item T5--T7 evaluate how these pairwise outcomes compose into a chain, fan-in, and a hidden shared ancestor. They are representative three-node compositions. 
    \item T8 is the combined stress case: an eight-node DAG with multiple inputs and outputs, a deeper chain, implication pruning, and two overlapping maximal cliques that must remain separate latent explanations.
\end{itemize}

In T1--T7, each observed endpoint reaches at most four terminal cache domains. T8 uses $12$ isolated domains; its deepest endpoints can reach ten of them, while the smallest terminal routing
probability is $3/35\approx8.57\%$, still above the $5\%$ resolution used by \framework{}. Every relay also has its own private cache domain, ensuring that shared and private cache reach can be distinguished.

\noindent\textbf{Results.}
Fig.~\ref{fig:suite} compares each configured topology with the topology recovered by \framework{}. Appendix~\ref{app:controlled-evidence} reports the per-direction hit/miss counts.
The results show that, \framework{} correctly recovers all eight configured topologies.
For T1--T4, it distinguishes all four pairwise relation classes. In T5, it
recovers the two adjacent containment relations and correctly skips the
transitive pair through implication pruning. T6 recovers the fan-in structure,
where two endpoints are contained in a common containing endpoint. T7
correctly separates the hidden shared upstream of $B$ and $C$ from their
common downstream node $A$.
For T8, \framework{} recovers all six direct observed containment relations and both latent explanations, $S_1\to\{B,C,D\}$ and $S_2\to\{C,G\}$, while pruning eight long-range pairs whose relations are already implied by the recovered order.

In this controlled evaluation, each configured supply edge is intentionally
aligned with a strict cache-containment relation. Therefore, exact recovery of
the containment topology also recovers the configured supply graph in this
evaluation.

\begin{findingbox}
\noindent\textbf{Finding 8.}
\framework{} exactly recovers all $8/8$ configured topologies.
Across these cases, it correctly distinguishes all four pairwise relation
classes and preserves the intended multi-hop composition, fan-in structure,
pruning, and overlapping latent-sharing explanations.
\end{findingbox}

\section{Discussion}
\label{sec:discussion}

\paragraph{Conservative recovery of cache containment.}
The cache-containment structure recovered by \framework{} is an under-approximation: every reported relation is supported by directional cache measurements, whereas an unreported relation remains unknown rather than absent.
Genuine containment can remain unobserved for
many reasons. For example, cache domains carrying less than the measurement
resolution $\delta$ may not produce enough observations to be distinguished
reliably; routing may vary during a measurement window because of
temporary changes in load balancing, rate limits, health status, fallback, or
quota availability, causing repeated requests to sample a different routing
mixture from the one that normally serves the endpoint. These effects
remove evidence rather than provide evidence for an additional
relation. 

\paragraph{Cache-reach topology vs.\ supply chain.}
As discussed in \S\ref{sec:classify}, the recovered topology describes relations among the cache reaches observable
through reseller endpoints, not the supply graph that connects those
resellers. A containment chain is therefore
consistent with multi-level resale, but its length should not be interpreted as
the number of commercial intermediaries traversed by a request. Similarly, a
highly contained cache reach indicates concentration of observable cache
reach, but does not identify the operator, contract, or exact forwarding path
responsible for that concentration. The security significance of the
recovered topology is thus structural: it identifies where nominally
independent endpoints expose overlapping or nested cache reach, and hence where
a shared upstream component could create correlated confidentiality or
integrity exposure, without claiming that a particular supplier relationship
or compromise exists.

\vspace{-5pt}
\paragraph{Potential mitigations.}
A stronger design would make resellers forwarders rather than trust termination
points. One possibility is an \emph{end-to-end context protocol}, analogous to
an authenticated SSH or TLS session, established between the user's client and
the final serving backend through the reseller chain. The client would
authenticate the serving backend and encrypt the prompt context to its key; the
backend would decrypt the context only at the final hop and return an encrypted,
signed response. The signature should bind a fresh request identifier, the
selected model, the response, and any tool-call arguments, and the client
should verify it before releasing the response to an agent runtime.
Intermediate resellers could still route an opaque envelope and account for
usage through backend-signed receipts, but could neither read the prompt
context nor silently rewrite the returned content or agent actions. Realizing such
a protocol would require provider cooperation and careful support for dynamic
routing, streaming, caching, and billing; we leave its design and deployment to
future work.

\section{Related Work}
\label{sec:related}

\paragraph{Co-residency detection in the cloud.}
Cloud side channels established that customers can infer co-residency, exploit
the resulting cross-tenant boundary, and revisit placement after provider
hardening~\cite{ristenpart2009cloud,zhang2012crossvm,inci2015coresidency,
varadarajan2015placement}. We transfer this structure from hardware resources
and hypervisors to prefix caches and commercial
intermediaries; web dependency measurement motivates the resulting
concentration question~\cite{kashaf2020dependency}.

\paragraph{LLM supply-chain measurement.}
Audits cover reseller behavior, model substitution, pricing, and malicious
gateways~\cite{liu2026agent,gatescope2026,cai2025paying}; routers select efficient
backends~\cite{chen2023frugalgpt,ong2024routellm}; and cooperative fingerprints
identify models~\cite{xu2024instructionalfp}. We instead infer the hidden dependencies of endpoints.

\paragraph{Prefix-cache and inference side channels.}
Prefix-cache attacks recover content, detect cross-user reuse, and motivate
partitioning~\cite{gu2025auditing,zheng2024inputsnatch,song2025earlybird,
luo2026shadow,pang2025cachepart}; broader channels reveal architecture, identity,
or activity~\cite{dutta2023spygpu,gao2025iknow,carlini2024stealinglm,
duddu2018stealing,hu2020deepsniffer,tramer2016stealing,
naghibijouybari2018rendered,mcdonald2025whisperleak,soleimani2025wiretapping,
zhang2025netecho}. We repurpose authentic cache reuse as a relational
infrastructure signal rather than a content leak.

\paragraph{Network tomography.}
Network tomography infers hidden topology from end-to-end
observations~\cite{vardi1996tomography,castro2004tomography,caceres1999multicast,
bu2002general,rabbat2006multisource,eriksson2010topology}. Our planted prefix
state replaces packet-path correlation but retains its identifiability limit:
external traces determine only an equivalence class without extra assumptions or
vantage points (Remark~\ref{rem:supply}, Appendix~\ref{app:shadow}).

\section{Conclusion}
\label{sec:conclusion}

This paper studies hidden dependencies created by multi-level resale in the LLM API reseller ecosystem. We present \framework{}, an API-only measurement method that exploits prefix-cache reuse to reveal shared cache reach and containment among reseller endpoints. 
Our measurements on $39$ real-world resellers uncover widespread sharing, deep dependency structures, and strong concentration. 
These findings show that seemingly independent resellers can rely on common hidden upstreams, creating ecosystem-level confidentiality and integrity risks with potentially large blast radii. Our work provides a foundation for more transparent and supply-chain-aware security assessment of LLM API services.

\section*{Ethics Considerations}
\label{sec:ethics}

The measurement used paid credentials on commercial endpoints and issued only
synthetic prompts: random padding plus a fresh nonce, containing no user data
and no content that could be mistaken for a real workload. It neither attempted
nor required the recovery of any other customer's prompt, and the oracle we use
reveals only whether \emph{our own} prompt was cached.

\paragraph{Authorization.}
Every request was issued under accounts the authors registered and paid for, at
published rates, through the documented chat-completion interface. We did not
bypass authentication, evade rate limits, exploit a software defect, or access
any resource an ordinary customer of the same endpoint could not access. What the
study does is send a legitimate customer's own prompts at a legitimate customer's
own rate and read the usage field the endpoint chooses to return. We nonetheless
treat the aggregate as a measurement study rather than routine usage, which is
why the load bounds below and the disclosure policy exist.

\paragraph{Load on third parties.}
Flooding places state in shared caches and consumes capacity, so we bounded it.
Safety caps stopped non-converging floods at $1000$ requests; sustained errors
quarantined an endpoint. Aggregate load was $1.1\times10^{6}$ requests over
$7.2$ hours across $39$ endpoints, about one request per second per endpoint,
which is within ordinary customer usage for the paid tiers involved. Keep-warm
traffic ran at a low fixed rate and was stopped as soon as the corresponding
probe direction finished. We paid for all traffic at published rates.

\paragraph{Disclosure policy.}
We withhold endpoint identities in this paper. This is a standing policy, fixed
before analysis and independent of what the measurement found: a reported
containment relation is a statement about infrastructure during one window and
for one model, and publishing it under a company name invites reading it as an
allegation about a commercial arrangement we did not observe. Appendix~\ref{app:endpoints}
gives each node's endpoints in redacted form, which preserves the shape of the
measurement without naming an operator.

\bibliographystyle{IEEEtran}
\bibliography{references}

\appendices

\section{Measurement Budget and Request Complexity} \label{app:measurement-budget} \S\ref{sec:build} describes how \framework{} assembles pairwise verdicts into a containment order; this appendix analyzes the request cost of that procedure. Let $N=|\mathcal{L}|$ and let $B_f=1000$ denote the per-direction flood cap. Profiling requires one request per endpoint. Each invocation of \Determine\ tests two directions (lines~\ref{line:det-dirab}--\ref{line:det-dirba}), and each direction uses at most $B_f$ flood requests (line~\ref{line:flood-terminate}) plus $\Nprove$ successful \Prove\ observations (line~\ref{line:prove-budget}), excluding keep-warm traffic and failed requests. Without implication pruning, \Build\ invokes \Determine\ for at most $\binom{N}{2}$ endpoint pairs, giving a worst-case foreground request complexity of \[ O\left(N+N^2(B_f+\Nprove)\right). \] When short-range relations are discovered first in a containment chain, the $N-1$ adjacent relations can imply all longer-range pairs. The number of direct pair measurements then reduces to $O(N)$, giving foreground complexity \[ O\left(N(B_f+\Nprove)\right). \] The realized request cost depends on how many candidate pairs are pruned and is reported empirically in \S\ref{sec:eval}.

\section{Redacted Endpoint Index}
\label{app:endpoints}

\S\ref{sec:eval:selection} describes how the measured reseller endpoints were
selected.
Table~\ref{tab:endpoints} lists the $34$ nodes of Fig.~\ref{fig:field} with the
commercial endpoints behind each one. Every domain is redacted to its public
suffix and the last character of its registrable label; the remaining
characters are replaced by one block each, and subdomain labels are dropped.
The redaction preserves what a reader needs to sanity-check the measurement,
namely how many distinct operators a node contracts and how their domains are
shaped, without naming any of them (\S\ref{sec:ethics}). Rows with more than
one entry are clone equivalence classes.

\begin{table*}[t]
\centering
\caption{The $34$ nodes of Fig.~\ref{fig:field}, their layer, and the
redacted endpoints they stand for. Each block replaces one character of
the registrable domain label.}
\label{tab:endpoints}
\footnotesize
\setlength{\tabcolsep}{5pt}
\renewcommand{\arraystretch}{1.15}
\begin{tabular}{@{}ccp{0.21\textwidth}ccp{0.21\textwidth}@{}}
\toprule
Node ID & Layer & Redacted Endpoints & Node ID & Layer & Redacted Endpoints \\
\midrule
E1 & 1 & \texttt{\rb{}\rb{}\rb{}\rb{}i.com} & E18 & 3 & \texttt{\rb{}\rb{}\rb{}\rb{}\rb{}\rb{}c.com} \\
E2 & 1 & \texttt{\rb{}\rb{}c.ai} & E19 & 3 & \texttt{\rb{}\rb{}\rb{}\rb{}\rb{}i.ai} \\
E3 & 2 & \texttt{\rb{}\rb{}\rb{}\rb{}\rb{}\rb{}\rb{}u.com} & E20 & 3 & \texttt{\rb{}\rb{}\rb{}\rb{}1.online} \\
E4 & 2 & \texttt{\rb{}\rb{}\rb{}\rb{}\rb{}o.com} & E21 & 4 & \texttt{\rb{}\rb{}\rb{}\rb{}\rb{}\rb{}\rb{}k.com} \newline \texttt{\rb{}\rb{}\rb{}\rb{}\rb{}\rb{}\rb{}i.com} \\
E5 & 2 & \texttt{\rb{}\rb{}\rb{}\rb{}\rb{}i.top} & E22 & 4 & \texttt{\rb{}\rb{}\rb{}\rb{}\rb{}\rb{}\rb{}\rb{}\rb{}\rb{}\rb{}l.com} \\
E6 & 2 & \texttt{\rb{}\rb{}\rb{}\rb{}\rb{}i.com} & E23 & 4 & \texttt{\rb{}\rb{}\rb{}\rb{}\rb{}\rb{}\rb{}\rb{}\rb{}\rb{}\rb{}e.ai} \\
E7 & 2 & \texttt{\rb{}\rb{}y.app} & E24 & 5 & \texttt{\rb{}\rb{}\rb{}\rb{}i.fun} \\
E8 & 2 & \texttt{\rb{}\rb{}\rb{}\rb{}\rb{}\rb{}\rb{}e.cn} \newline \texttt{\rb{}\rb{}\rb{}\rb{}\rb{}\rb{}\rb{}s.top} & E25 & 5 & \texttt{\rb{}\rb{}\rb{}\rb{}\rb{}\rb{}\rb{}\rb{}l.cn} \\
E9 & 2 & \texttt{\rb{}\rb{}\rb{}\rb{}\rb{}\rb{}t.ai} & E26 & 5 & \texttt{\rb{}\rb{}\rb{}\rb{}\rb{}\rb{}\rb{}\rb{}\rb{}\rb{}\rb{}i.top} \\
E10 & 2 & \texttt{\rb{}\rb{}\rb{}j.org} & E27 & 5 & \texttt{\rb{}\rb{}t.ge} \\
E11 & 2 & \texttt{\rb{}\rb{}\rb{}\rb{}\rb{}e.ai} & E28 & 5 & \texttt{\rb{}\rb{}\rb{}\rb{}\rb{}\rb{}\rb{}\rb{}\rb{}\rb{}\rb{}i.top} \\
E12 & 2 & \texttt{\rb{}\rb{}\rb{}\rb{}\rb{}\rb{}\rb{}i.com} & E29 & 6 & \texttt{\rb{}\rb{}\rb{}\rb{}\rb{}\rb{}\rb{}\rb{}a.com} \\
E13 & 2 & \texttt{\rb{}\rb{}\rb{}\rb{}i.com} & E30 & 6 & \texttt{\rb{}\rb{}\rb{}i.host} \\
E14 & 2 & \texttt{\rb{}\rb{}\rb{}\rb{}\rb{}e.vip} \newline \texttt{\rb{}\rb{}\rb{}\rb{}\rb{}i.top} \newline \texttt{\rb{}\rb{}\rb{}\rb{}\rb{}\rb{}\rb{}i.buzz} & E31 & 6 & \texttt{\rb{}\rb{}\rb{}\rb{}u.ai} \\
E15 & 3 & \texttt{\rb{}\rb{}\rb{}\rb{}\rb{}i.com} & E32 & 6 & \texttt{\rb{}\rb{}\rb{}\rb{}\rb{}\rb{}i.com} \\
E16 & 3 & \texttt{\rb{}\rb{}\rb{}\rb{}\rb{}u.net} & E33 & 6 & \texttt{\rb{}\rb{}\rb{}\rb{}\rb{}\rb{}i.com} \\
E17 & 3 & \texttt{\rb{}\rb{}\rb{}\rb{}3.one} & E34 & 7 & \texttt{\rb{}\rb{}\rb{}\rb{}i.com} \newline \texttt{\rb{}\rb{}\rb{}\rb{}m.link} \\
\bottomrule
\end{tabular}
\end{table*}

\section{Directional Evidence for Controlled Evaluation}
\label{app:controlled-evidence}

\S\ref{sec:eval:controlled} evaluates \framework{} on eight configured
topologies whose ground truth is known by construction.
Table~\ref{tab:t1t7evidence} reports the per-direction hit/miss
counts underlying those results.
Each row shows the observations from both \Flood--\Prove~directions
and the resulting relation recovered by \Classify.

\begin{table*}[t]
  \centering
  \caption{Directional evidence for T1--T7.
  $H/M$ denotes the numbers of hit and miss observations among
  successful probes.
  $X\rightsquigarrow Y$ floods $X$ and probes $Y$.}
  \label{tab:t1t7evidence}
  \small
  \setlength{\tabcolsep}{4.5pt}
  \resizebox{\columnwidth}{!}{\begin{tabular}{@{}cclll@{}}
    \toprule
    Topology & Endpoint pair & Forward $H/M$ & Reverse $H/M$ & Recovered relation \\
    \midrule
    T1 & $A,B$ & $A\rightsquigarrow B$: $0/59$ & $B\rightsquigarrow A$: $0/59$ & No relation \\
    T2 & $A,B$ & $A\rightsquigarrow B$: $179/2$ & $B\rightsquigarrow A$: $181/0$ & Equivalent \\
    T3 & $A,B$ & $A\rightsquigarrow B$: $180/1$ & $B\rightsquigarrow A$: $31/15$ & $B\to A$ \\
    T4 & $A,B$ & $A\rightsquigarrow B$: $9/15$ & $B\rightsquigarrow A$: $14/15$ & Latent $S\to\{A,B\}$ \\
    T5 & $A,B$ & $A\rightsquigarrow B$: $180/1$ & $B\rightsquigarrow A$: $42/15$ & $B\to A$ \\
       & $B,C$ & $B\rightsquigarrow C$: $181/0$ & $C\rightsquigarrow B$: $13/15$ & $C\to B$ \\
    T6 & $A,B$ & $A\rightsquigarrow B$: $181/0$ & $B\rightsquigarrow A$: $5/15$ & $B\to A$ \\
       & $B,C$ & $B\rightsquigarrow C$: $0/59$ & $C\rightsquigarrow B$: $0/59$ & No relation \\
       & $A,C$ & $A\rightsquigarrow C$: $181/0$ & $C\rightsquigarrow A$: $8/15$ & $C\to A$ \\
    T7 & $A,B$ & $A\rightsquigarrow B$: $180/1$ & $B\rightsquigarrow A$: $18/15$ & $B\to A$ \\
       & $B,C$ & $B\rightsquigarrow C$: $8/15$ & $C\rightsquigarrow B$: $8/15$ & Latent $S\to\{B,C\}$ \\
       & $A,C$ & $A\rightsquigarrow C$: $180/1$ & $C\rightsquigarrow A$: $20/15$ & $C\to A$ \\
    \bottomrule
  \end{tabular}}
\end{table*}

\section{Cross-Model Containment Relations}
\label{app:gpt51}

\begin{figure}[t]
\centering
\includegraphics[width=0.5\columnwidth]{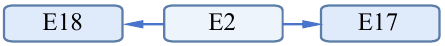}
\caption{The two cache-containment relations reproduced on both
\texttt{gpt-5-nano-2025-08-07} and \texttt{gpt-5.1}.
Arrows denote cache-reach containment, as in
Fig.~\ref{fig:field}.}
\label{fig:cross-model}
\end{figure}

\S\ref{sec:eval:sku} evaluates whether the cache-containment
relations recovered on \texttt{gpt-5-nano-2025-08-07} persist when the
same endpoint pairs are measured on \texttt{gpt-5.1}. Among the decisive
reruns, only two reproduce the original containment relation.
Figure~\ref{fig:cross-model} shows these two cases.

\section{Interpreting Latent Witnesses}
\label{app:shadow}

\S\ref{sec:build} introduces latent witnesses as part of \Build; this appendix
establishes that they are not identifiable.
\Build\ may encounter \textsc{Partial} pairs whose shared cache reach cannot be
explained by any observed endpoint already present in the recovered containment
structure. To represent such unexplained sharing without attributing it to a
specific reseller, \textsc{ShadowInsert} introduces latent witnesses. These
witnesses are explanatory placeholders only: they are never queried and should
not be interpreted as identified upstream services or commercial entities.

\paragraph{Why latent witnesses are not identifiable.}
The observations available to \framework{} are insufficient to determine the
internal structure behind a latent witness. For example, a hidden node that
explains a set of visible cache-reach relations could be replaced by a longer
forwarding chain, by several co-tenant API surfaces with equivalent cache
behavior, or by a combination of the two. As long as the replacement preserves
the cache-reach relations observed among the visible endpoints, all
\Profile, \Flood, \Prove, and \Determine\ outcomes remain unchanged because
every measurement begins and ends at a visible endpoint.

Consequently, the trace does not determine how many hidden forwarding nodes
exist, how they are internally connected, or whether two latent witnesses
correspond to the same underlying system. We therefore use a latent witness only
to record that some observed sharing remains unexplained by the visible
containment structure.

\begin{algorithm}[t]
\small
\caption{$\textsc{ShadowInsert}(R,\mathcal{P})$: overlapping latent witnesses for
unexplained sharing relations.}
\label{alg:shadow-insert}
\begin{algorithmic}[1]
\Require recovered reachability $R$, partial-pair set $\mathcal{P}$
\State $H\gets(V,\emptyset)$
\For{each $(u,v)\in\mathcal{P}$}
  \If{$\neg\textsc{ExplainedByObservedAncestor}(R,u,v)$}
    \State add $\{u,v\}$ to $H$
  \EndIf
\EndFor
\State $\mathcal{C}\gets\textsc{AllMaximalCliques}(H)$ of size $\ge2$
\For{each $C\in\mathcal{C}$ in lexicographic order}
  \State $s_C\gets\textsc{NewWitness}(C)$;
         $R[s_C][v]\gets1$ for each $v\in C$
\EndFor
\State \Return $(R,\{s_C\})$
\end{algorithmic}
\end{algorithm}

\paragraph{Representing unexplained sharing.}
Algorithm~\ref{alg:shadow-insert} constructs an auxiliary graph over unexplained
\textsc{Partial} pairs and inserts one latent witness for each maximal clique.
Using maximal cliques rather than greedily merging endpoints preserves
overlapping explanations: for example, T8 contains both $\{B,C,D\}$ and
$\{C,G\}$, which share endpoint $C$ but should remain distinct structural
explanations. A size-two clique simply represents a local pairwise witness; in
the field data, all 18 unexplained sharing relations are of this form.

Because latent witnesses are not identifiable from the measurement trace,
neither their number nor their labels should be interpreted as the number or
identity of hidden commercial entities. Accordingly, latent witnesses are
reported separately and are excluded from all observed-node counts,
containment-edge counts, and cache-reach statistics in
\S\ref{sec:eval}.

\end{document}